\documentclass[12pt,preprint]{aastex}
\begin{document}

\title{The FWHM of local pulses and the corresponding power-law index of gamma-ray burst FRED pulses}

\author{L.-W. Jia$^{1,3}$, Y.-P. Qin$^{1,2,4}$}
\affil{$^1$National Astronomical Observatories/Yunnan Observatory,
Chinese Academy of Sciences,\\P. O. Box 110, Kunming 650011, China\\
$^2$Physics science and technology institute, Guangxi University, Nanning, Guangxi 530004, P.R. China\\
$^3$The Graduate School of the Chinese Academy of Sciences\\
$^4$E-mail: ypqin@ynao.ac.cn; lwjia@ynao.ac.cn}

\begin{abstract}
The FWHM of gamma-ray burst (GRB) pulses is known to be related
with energy by a power-law. We wonder if the power-law index
$\alpha$ is related with the corresponding local pulse width
$FWHM_0$. Seven FRED (fast rise and exponential decay) pulse GRBs
are employed to study this issue, where six of them were
interpreted recently by the relativistic curvature effect (the
Doppler effect of fireballs) and the corresponding local pulses
were intensely studied. A regression analysis shows an
anti-correlation between $log \alpha$ and $log FWHM_0$ with a
slope of $-0.37\pm0.13$. This suggests that, for the class of the
GRB pulses which are consequences of the curvature effect, the
difference of the local pulse width might lead to the variation of
the power law index, where the smaller the width the larger the
value of $\alpha$. Since the number of sources employed in this
analysis is small, our result is only a preliminary one which
needs to be confirmed by larger samples.
\end{abstract}

\section{Introduction}

Gamma-ray bursts (GRBs) were discovered thirty-eight years ago by
chance, and now are recognized as the most luminous known objects
in the Universe (Fishman 2001). Since then, many observations of
the objects have been made, which have amassed a great deal of
information. Owing to their brief and random appearance in the
gamma-ray region, their study had become very difficult since
their discovery. Although the progress has been made in GRB
research, GRBs remain one of the most inexplicable astrophysical
phenomena observed today (Kouveliotou 1997).

Temporal and spectral characteristics of prompt emission of
gamma-ray burst pulses have been intensely studied since they
could constrain the energizing and emission mechanisms (Ryde et
al. 2002). The correlation between GRB spectral and temporal
properties have been investigated by several research groups. It
was first noted by Norris et al. (1986) that GRB pulses exhibit a
hard-to-soft spectral evolution, and associated with it the pulses
are seen to be narrower at higher energies than they are at lower
bands which were confirmed by later works (see Fishman et al.
1992; Link, Epstein, \& Priedhorsky 1993). By using the average
autocorrelation function and the average pulse width, Fenimore et
al. (1995) showed that the average pulse width has a power-law
dependence on energy with an index of about -0.4 (the range of it
is from -0.37 to -0.46, depending on how it is measured). This is
the first quantitative relationship between temporal and spectral
structure in gamma-ray bursts. Norris et al. (1996) found that
average raw pulse shape dependence on energy is approximately
power law, with an index of -0.40, consistent with the
autocorrelation analysis of Fenimore et al. (1995). Furthermore,
Nemiroff (2000) brought forward that over the energy range 100
keV-1 MeV in GRB 930214c (BATSE trigger 2193) the temporal scale
factors between a pulse measured at different energies are related
to that energy by a power law. The corresponding power-law indexes
found by Feroci et al. (2001) for GRB 990704 and by Piro et al.
(1998) for GRB 960720 are $-0.45$ and $-0.46\pm0.10$,
respectively. Costa (1999) also found that the power-law index for
GRB 960720 is $-0.46$, the same to Piro et al. (1998). The
spectral lag as a function of energy was examined for individual
pulses in GRBs (Norris et al. 2000), which confirmed the earlier
result of Fenimore et al. (1995). In a recent study (Crew et al.
2003), the power-law relationship between the duration of GRB
021211 and energy further confirmed the earlier result. The
anti-correlation between pulse widths and gamma-ray energy have
also been investigated by many other authors (e.g., Tavani 1997;
Wang et al. 2000; Beloborodov et al. 2000; Guidorzi et al. 2003;
Sakamoto et al. 2004; Dado et al. 2004).

Norris et al. (1995) found an anti-correlation between $T_{90}$
and peak intensity, while a positive correlation between $T_{90}$
and total fluence was shown in Lee \& Petrosian (1997), and a
positive correlation between peak energy and variability was found
by Lloyd-Ronning \&\ Ramirez-Ruiz (2002). A correlation between
luminosity and variability for BATSE bursts with known redshifts
was revealed by Fenimore \& Ramirez-Ruiz (2000). Ramirez-Ruiz \&
Fenimore (2000) found a quantitative relationship between pulse
amplitude and pulse width: the smaller amplitude peaks tend to be
wider, with the pulse width following a power law with an index of
about -2.8 (the range of it is from -2.8 to -3.0, depending on how
it is measured). The anti-correlation between the pulse amplitude
and pulse width was also revealed by Lee et al. (2000). $T_{90}$
being correlated with peak heights (Lestrade 1994) and peak energy
being correlated with peak flux (Mallozzi et al. 1995) were other
reported relationships.

It was suggested that, the most likely radiation progress in GRBs
is synchrotron emission in the standard fireball scenario (see
Katz 1994; Sari, Narayan, \& Piran 1996). The power-law dependence
has led to the suggestion that this effect could be attributed to
synchrotron radiation (see Piran 1999). Kazanas, Titarchuk, \& Hua
(1998) proposed that synchrotron cooling could well account for
the effect (see also Chiang 1998; Dermer 1998; and Wang et al.
2000). Fenimore et al. (1995) showed that synchrotron emission can
give rise to the correlation $t_{syn}(E)\propto E^{-0.5}$ between
GRB spectral and temporal properties, which is consistent with the
observed correlation $\Delta \tau \propto E^{-0.45\pm 0.05}$.
Cohen et al. (1997) put forward that the power-law relationship
between pulse width and energy with the index of $-0.4$ is in
reasonable agreement with expectations for a population of
electrons losing energy by synchrotron radiation, for which an
exponent of $-1/2$ is predicted. It was suspected that a simple
relativistic mechanism might be at work in producing this
relationship (Nemiroff 2000). In deed, it was shown recently in
Qin et al. (2004; hereafter Paper I) and Qin et al. (2005) that
the Doppler effect of a relativistically expanding fireball
surface (the so-called relativistic curvature effect) could lead
to a power law relationship between the pulse width and energy for
FRED (fast rise and exponential decay) pulses, regardless the real
forms of the rest frame radiation and the local (or intrinsic)
pulse involved. The same effect was also observed by Shen et al.
(2005).

In this paper, we investigate if local pulses are related with the
power law relationship (in other words, we wonder how the local
pulse width is related with the index of the power law observed).
In section 2, we choose several GRBs with each of them comprising
a single FRED pulse to calculate the corresponding data.
Relationship between the index and the local pulse width is
explored in section 3. Conclusions are presented in section 4.

\section{Sources and data}

To study how local pulses affect the index of the power law
between the pulse width and energy, we focus on FRED pulse bursts.
As revealed recently by many authors, the observed FRED structure
of pulses could be interpreted by the relativistic curvature
effect when the observed plasma moves relativistically towards us
and appears to be locally isotropic (see, e.g., Fenimore et al.
1996; Ryde \& Petrosian 2002; Kocevski et al. 2003; Paper I; Shen
et al. 2005). If this interpretation is correct, FRED pulses would
form in nature a class identified by the GRB temporal structure.
In this way, it would not be great surprise to us if quantities
associated with the pulses are correlated with each other.

As illustrated in Paper I, Shen et al. (2005), and Qin \& Lu
(2005), the local pulse width is essential to produce the observed
pulse shape due to the curvature effect. Accordingly, those FRED
pulses with their local pulses having been intensely studied
become our first choice. We find six bursts studied in Paper I
belonging to this kind. They are GRB 910721 ($\#563$), GRB 920925
($\#1956$), GRB 930612 ($\#2387$), GRB 941026 ($\#3257$), GRB
951019 ($\#3875$) and GRB 951102B ($\#3892$).

Light curve data for which the background counts have been
subtracted are available in the BATSE website
(http://cossc.gsfc.nasa.gov/batse/batseburst/sixtyfour\_ms\\/bckgnd%
\_fits.html). The signal data are taken within the zone [$t_{\min
},t_{\max } $], where $t_{\max }-t_{\min }=2T_{90}$, and $t_{\min
}$ is at $T_{90}/2$ previous to the start of $T_{90}$.

There might be many different methods to estimate the pulse width.
The theme of all possible methods is to find the central values of
the scattering data. In other words, one always manages to find
the real values of the data that are assumed to be get rid of the
chaos arising from the influence of the background as well as
other statistical errors. Owing to the fact that the light curve
function of Kocevski et al. (2003) (the KRL function; equation
[22] of Kocevski et al. 2003) could well describe the observed
light curves of FRED pulses (see also Qin \& Lu 2005), we simply
employ this function to fit the four channel light curves of the
six bursts, where parameters of the function associated with
different channels are allowed to be different for the same burst.
In order to allow the fitting curves shifting along the time axis
so that the time coordinate of the light curve data is unnecessary
to be resettled, we introduce an extra parameter $t_0$ to the KRL
function, where $t$ should be replaced by $t-t_0$ and $t_m$ should
be replaced by $t_m-t_0$. Thus, we have five free parameters
($f_m$, $t_m$, $r$, $d$, $t_0$) in our fit, instead of four. The
widths of the four channel light curves are then estimated from
the corresponding fitting curves, where the errors are determined
by the uncertainties of the fitting parameters via the error
transfer formula.

We perform the fit with the software of ORIGIN, where the fitting
parameters as well as their uncertainties are available.
Illustrated in Figure. 1 are the fits to the four channel light
curves of GRB 951019 ($\#3875$). For GRB 941026 ($\#3257$) and GRB
951102B ($\#3892$), the widths in channel 4 are not available
since the signal in that channel is too weak to be detected. The
estimated values of the FWHM of the observed light curves of the
six bursts calculated with the fitting curves (determined by the
fitting parameters) are listed in Table 1.

Assuming that the widths of pulses are related with energies by a
power law, we calculate the indexes with the estimated values of
the observed pulse widths of the six sources. The results are
presented in Table 2.

There are several local pulses discussed in Paper I, some of which
are as follows:

1. The local pulse with an exponential decay
\begin{equation}
\widetilde{I}(\tau _\theta )=I_0\exp (-\frac{\tau _\theta -\tau
_{\theta ,\min }}\sigma )\qquad(\tau _{\theta ,\min }\leq \tau
_\theta )
\end{equation}

2. The local pulse with a power-law rise and a power-law decay
\begin{equation}
\widetilde{I}(\tau _\theta )=I_0\{
\begin{array}{c}
(\frac{\tau _\theta -\tau _{\theta ,\min }}{\tau _{\theta ,0}-\tau
_{\theta ,\min }})^\mu \qquad(\tau _{\theta ,\min }\leq
\tau _\theta \leq \tau _{\theta ,0}) \\
(1-\frac{\tau _\theta -\tau _{\theta ,0}}{\tau _{\theta ,\max
}-\tau _{\theta ,0}})^\mu \qquad(\tau _{\theta ,0}<\tau _\theta
\leq \tau _{\theta ,\max })
\end{array}
\end{equation}

3. The local pulse with a power-law rise
\begin{equation}
\widetilde{I}(\tau _\theta )=I_0(\frac{\tau _\theta -\tau _{\theta ,\min }}{%
\tau _{\theta ,\max }-\tau _{\theta ,\min }})^\mu \qquad(\tau
_{\theta ,\min }\leq \tau _\theta \leq \tau _{\theta ,\max })
\end{equation}

In Paper I, local pulse (1) was adopted to account for the light
curves of GRB 910721 and GRB 930612 when the curvature effect was
considered, for GRB 941026 and GRB 951102B local pulse (2) with
$\mu =1$ was taken, while or GRB 920925 and GRB 951019 local pulse
(3) with $\mu =1$ was assumed. After smoothing the signal data,
they got a very good fit to these sources (see the $\chi^2$ values
listed in Table 2 of Paper I), which suggests that the assumption
that the observed light curves could arise from the local pulses
adopted when taking into account the curvature effect is
acceptable.

According to the local pulse parameters listed in Table 2 of Paper
I, we get from equations (1)-(3) the widths of the corresponding
local pulses (note that $\tau _{\theta ,\min }=0$ was adopted in
Paper I), which are listed in Table 2 in this paper as well.

\section{Relationship analysis}

Relation between the index of the power law, $\alpha$, and the
FWHM of the local pulses, $FWHM_0$, is displayed in Figure 2. A
linear correlation between $log \alpha$ and $log FWHM_0$ could be
observed.

We wonder if sources other than those selected in Paper I are in
agreement with this trend. As a FRED pulse source, GRB 930214c
($\#2193$) was previously intensely studied (Nemiroff 2000). We
include this burst in our study. As done in the case of the six
bursts, we once more employ the KRL function to fit the four
channel light curves of this source, and in the same way,
parameters of the function associated with different channels are
allowed to be different. Also, the fit is performed with the
software of ORIGIN. The widths of the 4 channels of this burst are
estimated with the fitting parameters, which are listed in Table 1
as well. From these widthes we get the power law index of this
source under the assumption that the widths are related with
energies by a power law. The estimated value of the index is
presented in Table 2.

To obtain the local pulse width of this burst, we follow what were
done by Qin et al. in Paper I for the six GRBs. One can find the
details of the analysis in the mentioned paper, which are omitted
in the following. Briefly stating, we fit the count rate of the
third channel of GRB 930214c ($\#2193$) with equation (21) of
Paper I, where local pulse (1) in this paper (which is local pulse
[83] in Paper I) is adopted. Relations (e.g., $t=t_1\tau+t_0$, see
Paper I for a detailed explanation) and functions (e.g., DB3) and
the corresponding parameters taken for the fit are exactly those
adopted in Paper I in the case of GRB 910721. The fit yields:
$\sigma=1.70$, $\chi^2_\nu=0.416$ for the data smoothed with DB3
wavelet in the level of the first-class decomposition,
$\chi^2_\nu=0.819$ for the data without smoothing, and other free
parameters (they are not related to the local pulse width). (Owing
to the limited space provided, the figure showing the fit is
omitted.) We find for GRB 930214c ($\#2193$) that the reduced
$\chi^2$ associated with the fit is reasonable, which suggests
that the light curve of this burst could indeed be accounted for
by the relativistic curvature effect.

The data point of ($\alpha$, $FWHM_0$) for GRB 930214c ($\#2193$)
is also plotted in Figure 2, which is in agreement with the trend
mentioned above (see Figure 2).

A linear correlation analysis of the data of the seven bursts
yields: $log \alpha=(-0.43\pm0.06)+(-0.38\pm0.11)log FWHM_0 $
($r=-0.84$, $N=7$). However, it should be noticed that the number
of the sources concerned is small. In this case, the result of the
correlation analysis might obviously depend on some lonely located
data points (see GRB 920925 and GRB 930214c in Figure 2).
According to Isobe et al. (1990) and Feigelson \& Babu (1992), the
true regression coefficient uncertainty in samples of small size
would be underestimated when the usual standard formulas are
applied. Thus, resampling procedures such as the jackknife or
bootstrap should be used to evaluate regression uncertainties in
these cases. We thus try to use the bootstrap method to estimate
the regression coefficient uncertainties. Applying the bootstrap
error analysis we get indeed a larger slope uncertainty: $log
\alpha=(-0.43\pm0.08)+(-0.37\pm0.13)log FWHM_0 $. This is what we
should hold.

\section{Conclusions}

In this paper, we investigate the relationship between the
power-law index $\alpha$ and the FWHM of local pulses, $FWHM_0$,
of seven FRED pulse GRBs. Our analysis shows that there exists a
linear relationship with a slope of $-0.37\pm0.13$ between $log
\alpha$ and $log FWHM_0$ for the bursts. This suggests that
different widths of local pulses could lead to different values of
the power law indexes, with the larger the former the smaller
absolute value the latter, for FRED pulse bursts (at least for
those which were previously interpreted by the relativistic
curvature effect). If this relationship could be confirmed, the
distribution of the local pulse width would be an important factor
that leads to the variation of the index observed in GRB samples
(this might likely be true if the sample contains only FRED
pulses).

Of the seven GRBs, local pulses of six were previously intensely
studied and that of the other one is explored in this paper. As
the number of the sources involved is small, the result is only a
preliminary one, which is not at all conclusive in terms of
statistics. However, a trend in the relationship is explicitly
illustrated in our analysis, although the analysis is qualitative
rather than quantitative. A large sample of FRED pulses is thus
required to check statistically if this conclusion could hold.

Say frankly, the cause of this relationship is currently unclear.
Since the seven bursts studied here are single FRED pulse sources
which were assumed to suffer from the relativistic curvature
effect, we suspect that it might be this effect that gives birth
to the relationship. A theoretical analysis on this issue is
necessary.

Besides the curvature effect, there might be other factors that
can affect the value of the power law index. One would be the
variation of the rest frame emission mechanism, which was revealed
in Qin et al. (2005). For example, different rest frame spectra or
different speeds of the rest frame spectral softening could lead
to different values of the power law index. This also requires a
further investigation.

\acknowledgments Our thanks are given to Dr. Robert Nemiroff for
providing us helpful suggestions which make the paper
significantly improved. This work was supported by the Special
Funds for Major State Basic Research Projects (``973'') and
National Natural Science Foundation of China (No. 10273019).

\clearpage

\begin{figure}[tbp]
\includegraphics[angle=0,scale=1.5]{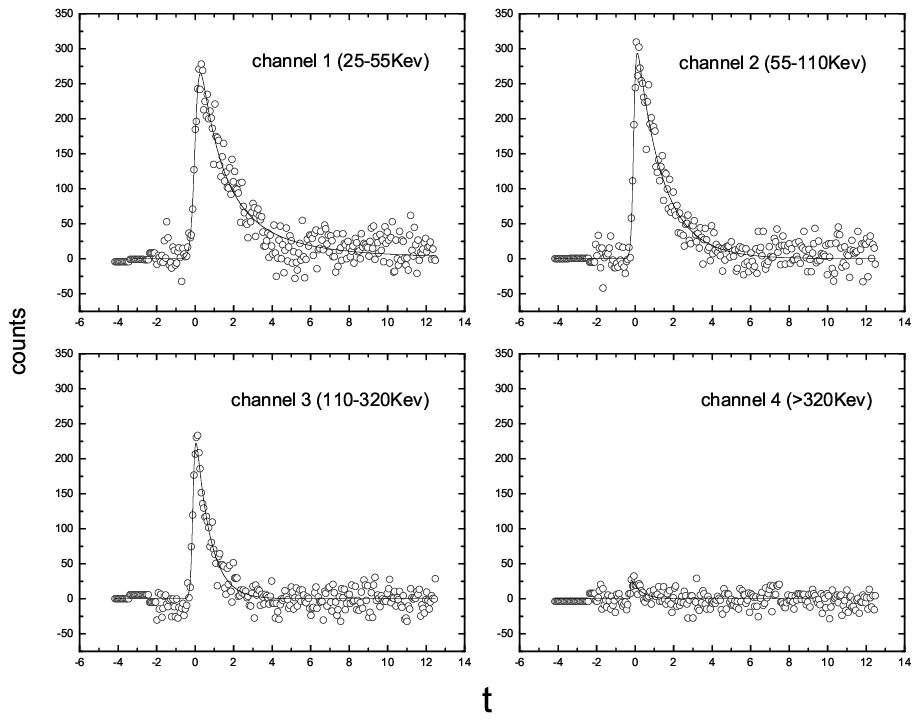}
\caption{Illustration of the fit with the KRL function to the four
channel light curves of GRB 951019.}
\end{figure}

\clearpage

\begin{figure}[tbp]
\includegraphics[angle=0,scale=1.5]{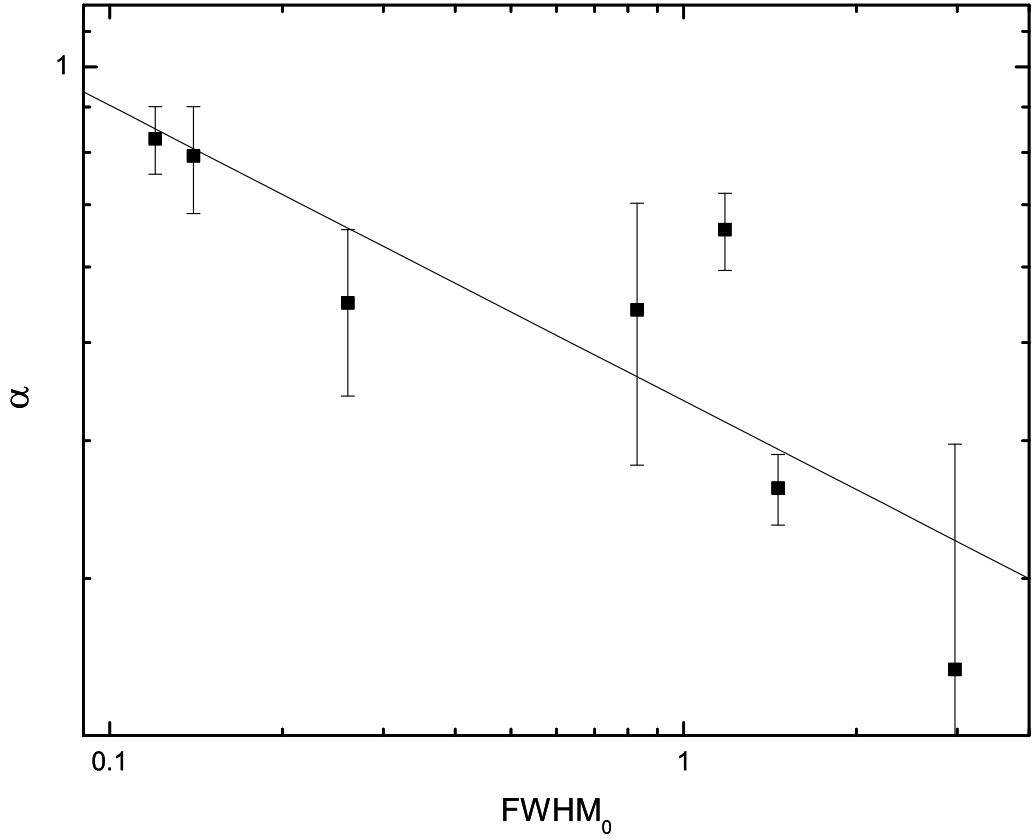}
\caption{Plot of the power law index versus the FWHM of local
pulses. The solid line is the linear regression line of the data.
}
\end{figure}

\clearpage

\begin{table*}
\begin{flushleft}

Table 1. The FWHM of the observed light curves of various channels
estimated with the KRL function for GRB 910721($\#563$), GRB
920925($\#1956$), GRB 930214c($\#2193$), GRB 930612($\#2387$), GRB
941026($\#3257$), GRB 951019($\#3875$) and GRB
951102B($\#3892$), respectively.\\[0pt]

\centering
 \vspace{0.2cm}
 \tabcolsep0.02in
\begin{tabular}[htb]{cccccccc} \hline \hline

GRB&trigger&&channel&&$FWHM$(s)&&$\sigma_{FWHM} $ \\
\hline
910721 &$\#$563  && 1 && 13.16 && 3.49 \\
       &     && 2 && 9.55  && 0.75 \\
       &     && 3 && 5.35  && 0.36 \\
       &     && 4 && 2.30  && 0.77 \\
\hline
920925 &$\#$1956 && 1 && 5.55  && 0.76 \\
       &     && 2 && 4.68  && 0.73 \\
       &     && 3 && 4.23  && 0.82 \\
       &     && 4 && 5.91  && 4.13 \\
\hline
930214c&$\#$2193 && 1 && 36.65 && 7.74 \\
       &     && 2 && 43.69 && 2.00 \\
       &     && 3 && 26.79 && 0.75 \\
       &     && 4 && 13.93  && 2.40 \\
\hline
930612 &$\#$2387 && 1 && 18.78 && 0.75 \\
       &     && 2 && 15.39 && 0.39 \\
       &     && 3 && 12.39 && 0.31 \\
       &     && 4 && 6.85  && 2.53 \\
\hline
941026 &$\#$3257 && 1 && 23.86 && 3.61 \\
       &     && 2 && 17.07 && 1.03 \\
       &     && 3 && 8.98  && 0.25 \\
\hline
951019 &$\#$3875 && 1 && 1.58  && 0.22 \\
       &     && 2 && 1.33  && 0.20 \\
       &     && 3 && 0.77  && 0.08 \\
       &     && 4 && 0.59  && 0.58 \\
\hline
951102B&$\#$3892 && 1 && 2.99  && 0.38 \\
       &     && 2 && 1.84  && 0.49 \\
       &     && 3 && 1.47  && 0.30 \\
\hline \hline
\end{tabular}
\end{flushleft}
\end{table*}

\clearpage

\begin{table*}[tbp]
\begin{flushleft}

Table 2. Estimated values of the power law index of the
7 bursts and the FWHM of the corresponding local pulses. \\[0pt]

\centering

\vspace{0.1cm}
 \tabcolsep0.02in
\begin{tabular}[htb]{cccccccccc} \hline \hline

GRB&trigger&&$\alpha$&&&&$\sigma_{\alpha} $&&$FWHM_0$ \\
\hline
910721 & $\#$563  && 0.77&&&& 0.12&& 0.14 \\
\hline
920925 & $\#$1956 && 0.17&&&& 0.16&& 2.97 \\
\hline
930214c & $\#$2193 && 0.62&&&& 0.07&& 1.18    \\
\hline
930612 & $\#$2387 && 0.29&&&& 0.03&&1.46 \\
\hline
941026 & $\#$3257 && 0.81&&&& 0.08&& 0.12 \\
\hline
951019 & $\#$3875 && 0.50&&&& 0.12&& 0.26 \\
\hline
951102B & $\#$3892 && 0.49&&&& 0.18&& 0.41 \\
\hline \hline
\end{tabular}
\end{flushleft}
\end{table*}

\label{lastpage}

\end{document}